\documentclass[conference,10pt]{IEEEtran}
\usepackage{amsmath}
\usepackage{amssymb}
\usepackage{graphicx}
\usepackage{latexsym}
\usepackage{subfigure,verbatim}
\usepackage{amsfonts}
\usepackage{paralist}
\usepackage{times,url}
\usepackage{graphics,multirow}
\usepackage{color}
\usepackage{wrapfig}
\usepackage{algorithmic}
\usepackage{algorithm,fancyhdr}

\makeatletter
\def\ps@headings{%
\def\@oddhead{\mbox{}\scriptsize\rightmark \hfil \thepage}%
\def\@evenhead{\scriptsize\thepage \hfil \leftmark\mbox{}}%
\def\@oddfoot{}%
\def\@evenfoot{}}
\makeatother
\usepackage[top=.75in, bottom=1.0in, left=.625in, right=.625in]{geometry}

\newcommand{\ignore}[1]{}

\begin{document}

\title{An Evasion and Counter-Evasion Study in Malicious Websites Detection}

\author{Li Xu$^\dagger$ \hspace{5mm} Zhenxin Zhan$^\dagger$ \hspace{5mm} Shouhuai Xu$^\dagger$ \hspace{5mm} Keying Ye$^\ddagger$\\
$^\dagger$Deptartment of Computer Science, University of Texas at San Antonio \\
$^\ddagger$Department of Management Science and Statistics, University of Texas at San Antonio
}

\maketitle

\begin{abstract}
Malicious websites are a major cyber attack vector, and effective detection of them is an important cyber defense task.
The main defense paradigm in this regard is that the defender uses some kind of machine learning algorithms to train a detection model,
which is then used to classify websites in question.
Unlike other settings, the following issue is inherent to the problem of malicious websites detection:
the attacker essentially has access to the same data that the defender uses to train his/her detection models.
This `symmetry' can be exploited by the attacker, at least in principle, to evade the defender's detection models.
In this paper, we present a framework for characterizing the evasion and counter-evasion interactions between the attacker and the defender,
where the attacker attempts to evade the defender's detection models by taking advantage of this symmetry.
Within this framework, we show that an adaptive attacker can make malicious websites evade powerful detection models, but
{\em proactive training} can be an effective counter-evasion defense mechanism.
The framework is geared toward the popular detection model of decision tree, but can be adapted to accommodate other classifiers.
\end{abstract}


\section{Introduction}

Compromising websites and abusing them to launch further attacks (e.g., drive-by-download \cite{js:www:10,provos08sec}) have become one of the mainstream attack vectors. 
Unfortunately, it is infeasible, if not impossible, to completely eliminate such attacks, meaning that
we must have competent solutions that can detect compromised/malicious websites as soon as possible. 
The {\em dynamic} approach, which is often based on client honeypots or variants, can detect malicious websites with
high accuracy, but is limited in terms of its scalability.
The {\em static} approach, which often analyzes the website contents and then uses some detection models (e.g., decision trees) to classify them into benign/malicious classes,
is very efficient, but suffers from its limited success in dealing with sophisticated attacks (e.g. JavaScript obfuscation).
This hints that there is perhaps some inherent limitation in the trade-off between scalability and detection effectiveness.

In this paper, we bring up another dimension of the problem, which may have a fundamental impact on the aforementioned inherent limitation.
Unlike in other settings, the following issue is inherent to the problem of malicious websites detection:
The attacker essentially has access to the same data that the defender uses to train its detection models.
This `symmetry' could be exploited by the attacker to evade the defender's detection models.
This is because the attacker can effectively train and obtain (almost) the same detection models,
and then exploit them to make other malicious websites evasive.
This is possible because the attacker can manipulate the contents of malicious websites 
during the course of compromising them,
or after they are compromised but before they are analyzed by the defender's detection models.
This is feasible because the attacker controls the malicious websites.

More specifically, we make two contributions.
First, we propose a framework for characterizing the evasion and counter-evasion interactions between the attacker and the defender.
The framework accommodates a set of adaptive attacks against a class of detection models known as decision trees \cite{c45:1993},
which have been widely used in this problem domain.
The framework also accommodates the novel idea of {\em proactive training} as the counter-evasion mechanism against the
adaptive attacks, where the defender proactively trains its detection models while taking adaptive attacks into consideration.
Although the framework is geared towards decision trees, it can be adapted to accommodate other kinds of classifiers.

Second, we use a dataset that was collected during the span of 40 days to evaluate the evasion power of adaptive attacks
and the counter-evasion effectiveness of proactive training.
Experimental results show that an adaptive attacker can make malicious websites evade powerful detection models, but
{\em proactive training} can be an effective counter-evasion defense mechanism.
In order to deepen our understanding of the evasion and counter-evasion interactions, 
we also investigate which features (or attributes) of websites have a high security significance, namely that
their manipulation causes the misclassification of malicious websites.
Surprisingly, we find that the features of high security significance
are almost different from the features that would be selected by the standard feature selection algorithms.
This suggests that we might need to design new machine learning algorithms to best fit the domain of security problems.
Moreover, we find that the detection accuracy of proactively-trained detection models 
increases with the degree of the defender's proactiveness (i.e. the number of training iterations).  
Finally, we find that if the defender does not know the attacker's adaptation strategy, the defender should adopt 
the ${\tt full}$ adaptation strategy that will be described later.

The rest of the paper is organized as follows. 
Section \ref{sec:data-collection} briefly reviews the context of the present study. 
Section \ref{sec:resilience} investigates the framework of evasion and counter-evasion interactions.
Section \ref{sec:evaluating-framework} evaluates the effectiveness of the framework.
Section \ref{sec:related-work} discusses related prior work.
Section \ref{sec:conclusion} concludes the paper.

\section{Preliminaries}
\label{sec:data-collection}

In order to illustrate the power of adaptive attacks and the effectiveness of our counter-measure against them, we need to consider some concrete detection scheme. 
Since J48 classifiers 
are known to be successful in detecting malicious websites
\cite{driveby:prophiler:www:2011,cantina:2011,codaspy13:li,Ludl:2007,dt:static:2008,li:ccs:2012},
we adopt the detection scheme we proposed in \cite{codaspy13:li} as the starting point of the present study.
We showed in \cite{codaspy13:li} that J48 classifier outperforms Naive Bayes, Logistic and SVM classifiers.

We also inherit the data collection method described in \cite{codaspy13:li}.
At a high level, a crawler is used to fetch the website content corresponding to an input URL, benign and malicious alike.
Each URL is described by 105 application-layer features 
%(including information that can be obtained through the Whois, Geographic and DNS services) 
and 19 network-layer features \cite{codaspy13:li}. 
%The exploitation of cross-layer information is actually a novel contribution of \cite{codaspy13:li}.
%We reuse the 105 application-layer features and 19 network-layer features we defined in \cite{codaspy13:li}, 
%which was showed to be effective in detecting malicious websites.
We now briefly review the following 16 features that will be encountered later: 
{\tt URL\_length} (length of URL);
{\tt Content\_length} (the content-length field in HTTP header, which may be manipulated by malicious websites);
{\tt \#Redirect} (number of redirects);
{\tt \#Scripts} (number of scripts);
{\tt \#Embedded\_URL} (number of URLs embedded);
{\tt \#Special\_character} (number of special characters in a URL);
{\tt \#Iframe} (number of iframes);
{\tt \#JS\_function} (number of JavaScript functions in a website);
{\tt \#Long\_string} (number of strings with 51 or more letters in embedded JavaScript programs);
{\tt \#Src\_app\_bytes} (number of bytes communicated from crawler to website);
{\tt \#Local\_app\_packet} (number of crawler-to-website IP packets, including redirects and DNS queries);
{\tt Dest\_app\_bytes} (volume of website-to-crawler communications);
{\tt Duration} (the time it takes for the crawler to fetch the contents of a website, including rediects);
{\tt \#Dist\_remote\_tcp\_port} and {\tt \#Dist\_remote\_IP} (number of distinct TCP ports and IP addresses the crawler uses to fetch websites contents, respectively);
{\tt \#DNS\_query} (number of DNS queries);
{\tt \#DNS\_answer} (number of DNS server's responses).

The main notations are summarized as follows.
\begin{center}
{\small
\begin{tabular}{|r|p{.34\textwidth}|}
\hline
${\sf MLA}$ & machine learning algorithm \\
${\sf fv}$ & feature vector representing a website\\
$X_z$ & feature $X_z$'s domain is $[\min_z,\max_z]$ \\
$M_0,\ldots,M_\gamma$ & defender's detection schemes (e.g., J48 classifier)\\
$D'_0$ & training data (feature vectors) for learning $M_0$ \\
$D_0$ & $D_0=D_0.malicious\cup D_0.benign$, where malicious feature vectors in $D_0.malicious$ may have been manipulated \\
$D^\dag_0$ & feature vectors used by defender to proactively train $M_1,\ldots,M_\gamma$; $D_0^\dag=D_0^\dag.malicious\cup D_0^\dag.benign$ \\
$\alpha$, $\gamma$  &  number of adaptation iterations\\
$M_i(D_\alpha)$     & applying detection scheme $M_i$ to classify feature vectors $D_\alpha$  \\
$M_{0\text{-}\gamma}(D_\alpha)$ & majority vote of $M_{0}(D_\alpha), \ldots, M_{\gamma}(D_\alpha)$ \\
${\sf ST}, {\sf C}, {\sf F}$    & adaptation strategy ${\sf ST}$, manipulation algorithm ${\sf F}$, manipulation constraints ${\sf C}$ \\
$s\stackrel{R}\gets S$ & assigning $s$ as a random member of set $S$ \\
\hline
\end{tabular}
}
\end{center}

\section{Evasion and Counter-Evasion Framework}
\label{sec:resilience}

Adaptive attacks are possible because an attacker can collect the same data
as what is used by the defender to train a detection scheme.
The attacker also knows the machine learning algorithm(s) the defender uses
or even the defender's detection scheme.
To accommodate the worst-case scenario, we assume there is a single attacker that coordinates the compromise of websites (possibly by many sub-attackers).
This means that the attacker knows which websites are malicious, while the defender aims to detect them.
In order to evade detection,
the attacker can manipulate some features of the malicious websites.
The manipulation operations can take place during the course of compromising websites,
or after compromising websites but before they are classified by the defender's detection scheme.

\subsection{Framework Overview}

\begin{figure}[!hbtp]
\centering
\includegraphics[height=0.2\textwidth,width=0.46\textwidth]{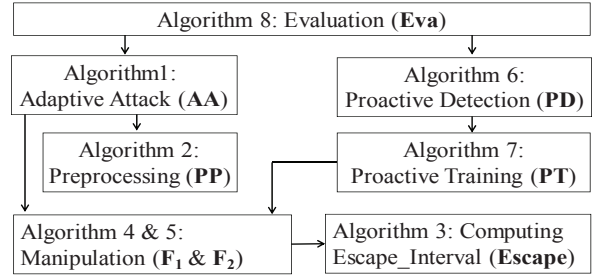}
\caption{Caller-callee relation between the algorithms.
\label{fig:calling-hierarchy}}
\end{figure}

We describe adaptive attacks and countermeasures in a modular fashion, by using eight algorithms
whose caller-callee relation is highlighted in Figure \ref{fig:calling-hierarchy}.
Algorithm \ref{resilience-master-attacker} is the attacker's main algorithm, which calls Algorithm \ref{preproc} for preprocessing,
and calls Algorithm \ref{resilience-alg1} or Algorithm \ref{resilience-alg2} for selecting features to manipulate and for determining the manipulated values for the selected features.
Both Algorithm \ref{resilience-alg1} and Algorithm \ref{resilience-alg2} call Algorithm \ref{alg:resilience-constraints}
to compute the {\em escape intervals} for the features that are to be manipulated.
An escape interval defines the interval from which the manipulated value of a feature should be taken so as to evade detection.

Algorithm \ref{protect-detectin-algorithm} is the defender's main algorithm, which calls Algorithm \ref{fig:pda-master} for proactive training of detection schemes.
For this purpose, the defender can have access to its own proactive manipulation algorithm.
In our experiments, we let the defender have access to the manipulations algorithms that are available to the attacker,
namely Algorithm \ref{resilience-alg1} and Algorithm \ref{resilience-alg2}.
This is sufficient for the purpose of understanding the effectiveness of proactive training and detection
against adaptive attacks under various algorithm/parameter possibilities, such as:
the defender correctly or incorrectly ``guess" the manipulation algorithm or parameters that are used by the attacker,
and the relatively more effective proactive training strategy against a class of adaptive attacks.
In order to evaluate the effectiveness of proactive training and detection against adaptive attacks,
we use an ``artificial" Algorithm \ref{protect-defense-vs-adaptive-attack-algorithm}, which is often implicit in most real-life defense operations.

\subsection{Evasion Model and Algorithms}

In our model, a website is represented by a feature vector.
We call the feature vector representing a benign website {\em benign feature vector},
and the feature vector representing a malicious website {\em malicious feature vector}.
Denote by $D'_0$ the defender's {\em training data},
namely a set of feature vectors corresponding to a set of benign websites (denoted by $D'_0.benign$) and malicious websites (denoted by $D'_0.malicious$).
The defender uses a machine learning algorithm {\sf MLA} to learn a detection scheme $M_{0}$ from $D'_{0}$ (i.e., $M_0$ is learned from one portion
of $D'_0$ and tested via the other portion of $D'_0$).
As mentioned above, the attacker is given $M_{0}$ to accommodate the worst-case scenario.
Denote by $D_0$ the set of feature vectors that are to be classified by $M_0$ to determine which feature vectors (i.e., the corresponding websites) are malicious.
The attacker's objective is to manipulate the malicious feature vectors in $D_0$
into some $D_\alpha$ so that $M_0(D_\alpha)$ has a high false-negative rate,
where $\alpha>0$ represents the number of iterations (or rounds) the attacker conducts the manipulation operations.

\begin{figure*}[!hbtp]
\centering
\subfigure[Parallel adaptation strategy]{\label{fig:attack-adaptation-methodologies-par}
\includegraphics[height=0.3\textwidth,width=0.29\textwidth]{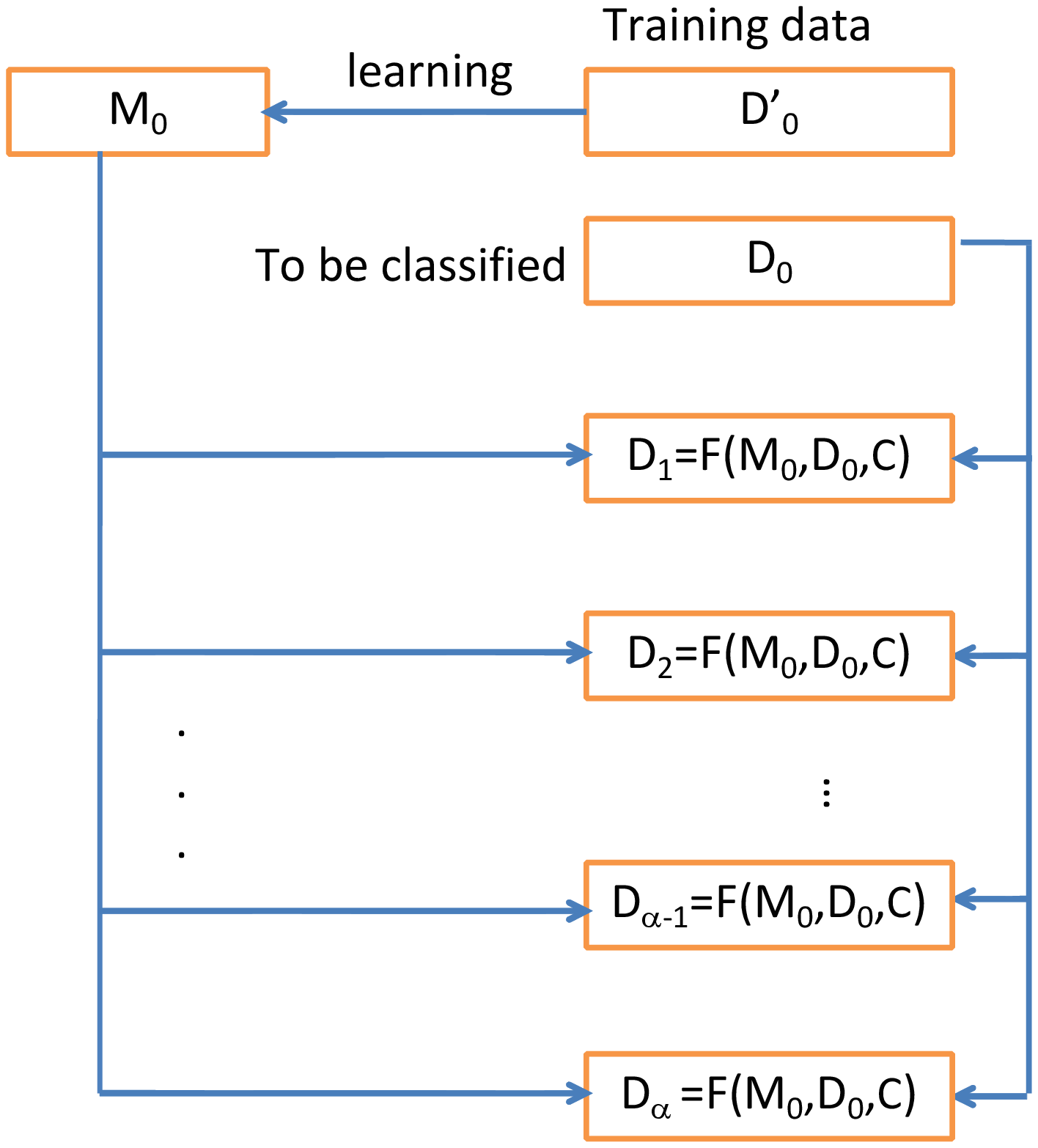}}
\hspace{2em}
\subfigure[Sequential adaptation strategy]{\label{fig:attack-adaptation-methodologies-seq}
\includegraphics[height=0.3\textwidth,width=0.29\textwidth]{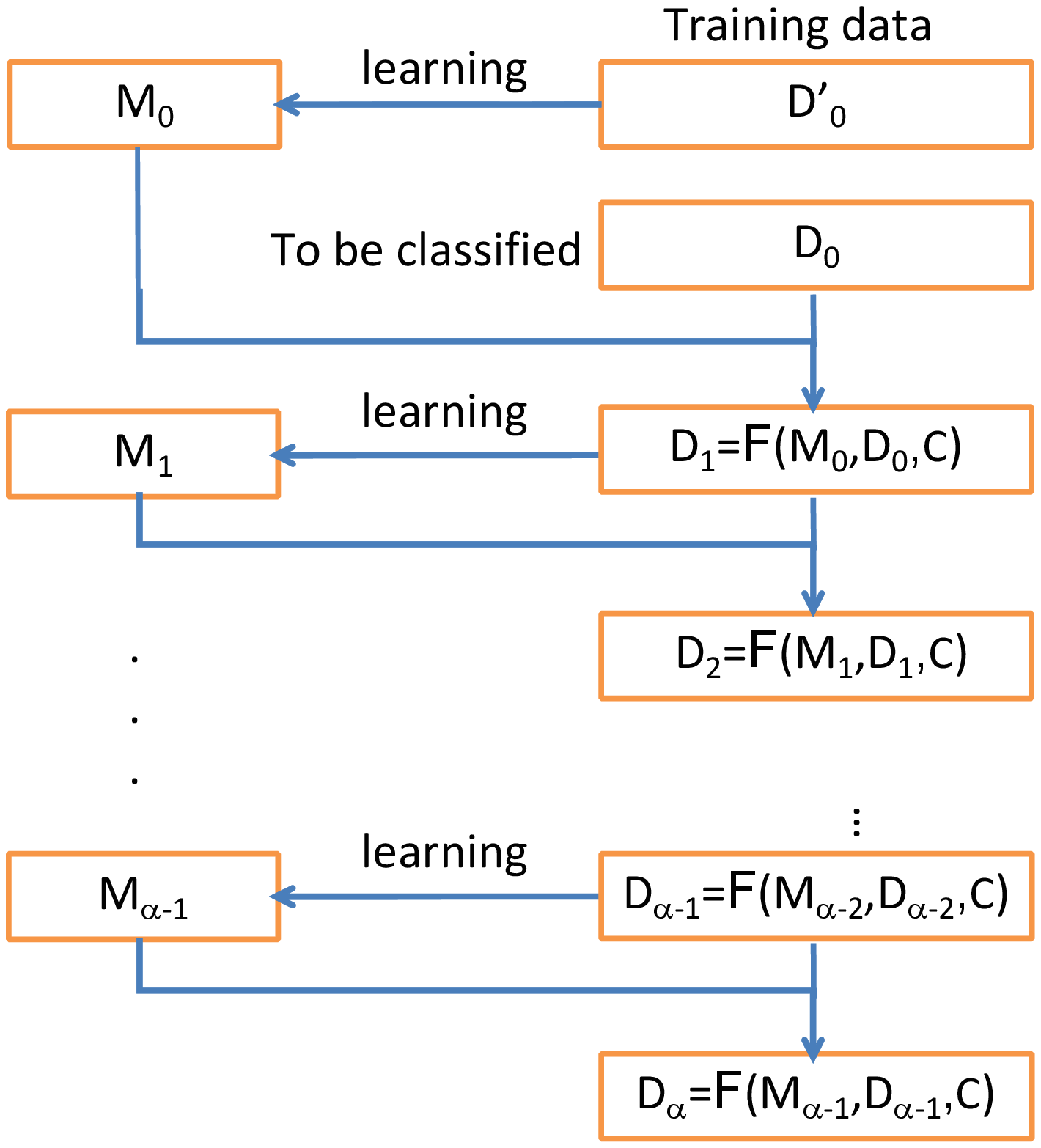}}
\hspace{2em}
\subfigure[Full adaptation strategy]{\label{fig:attack-adaptation-methodologies-full}
\includegraphics[height=0.3\textwidth,width=0.29\textwidth]{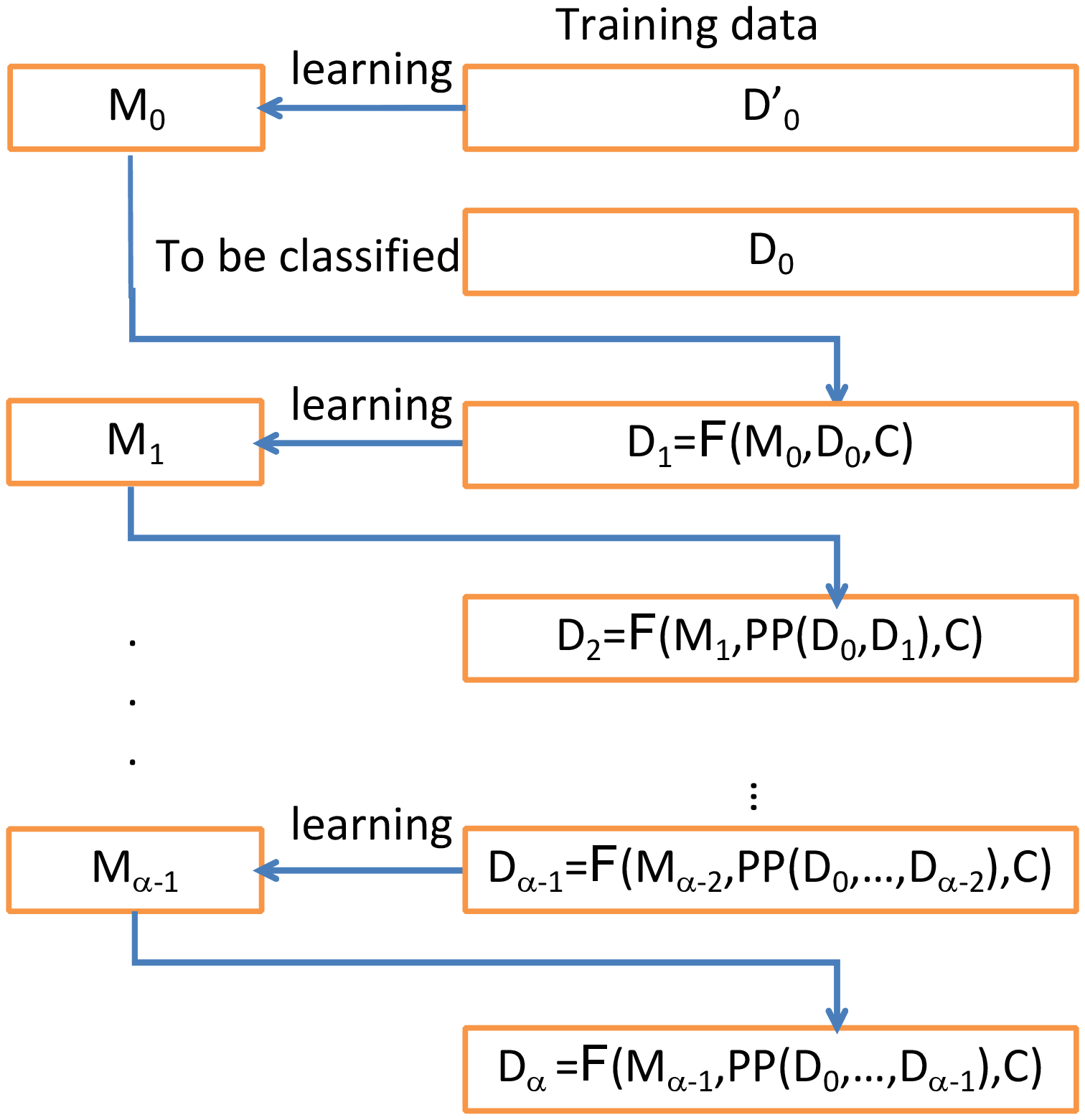}}
\caption{Adaptive attack algorithm ${\sf AA}({\sf MLA},M_0,D_0,{\sf ST},{\sf C}, {\sf F},\alpha)$,
where $D'_0$ is the defender's training data,
$M_0$ is the defender's detection scheme that is learned from $D'_0$ by using ${\sf MLA}$,
$D_0$ is the feature vectors that are examined by $M_0$ in the absence of adaptive attacks,
${\sf ST}$ is the attacker's adaptation strategy,
${\sf C}$ is a set of manipulation constraints,
${\sf F}$ is the attacker's (deterministic or randomized) manipulation algorithm that maintains the set of constraints {\sf C},
$\alpha$ is the number of rounds the attacker runs its manipulation algorithms.
$D_\alpha$ is the manipulated version of $D_0$ with malicious feature vectors $D_0.malicious$ manipulated.
The attacker's objective is make $M_0(D_\alpha)$ have high false-negative rate.}
\label{fig:attack-adaptation-methodologies}
\vspace{-0.2in}
\end{figure*}

\begin{algorithm}[hbt]
{\small
\caption{Adaptive attack ${\sf AA}({\sf MLA}, M_0, D_0, {\sf ST}, {\sf C},{\sf F}, \alpha)$}
\label{resilience-master-attacker}
INPUT:$M_0$ is defender's detection scheme,
$D_0=D_0.malicious\cup D_0.benign$ where malicious feature vectors ($D_0.malicious$) are to be manipulated (to evade detection of $M_0$),
{\sf ST} is attacker's adaptation strategy,
${\sf C}$ is a set of manipulation constraints,
{\sf F} is attacker's manipulation algorithm,
 $\alpha$ is attacker's number of adaptation rounds \\
OUTPUT: $D_\alpha$ %(replacement of $D_0$ in detection)
\begin{algorithmic}[1]
\STATE initialize array $D_1,\ldots,D_{\alpha}$
\FOR{$i$=1 $\TO$ $\alpha$}
\IF{{\sf ST} == {\tt parallel-adaptation}}
\STATE {$D_{i} \gets {\sf F}(M_{0},D_{0},{\sf C})$} ~~ \COMMENT{manipulated version of $D_{0}$}
\ELSIF{{\sf ST} == {\tt sequential-adaptation}}
   \STATE {$D_{i} \gets {\sf F}(M_{i-1},D_{i-1},{\sf C})$}~~\COMMENT{manipulated version of $D_{0}$}
\ELSIF{{\sf ST} == {\tt full-adaptation}}
     \STATE {${\cal D}_{i-1} \gets {\sf PP}(D_0,\ldots,D_{i-2})$}~~\COMMENT{see Algorithm \ref{preproc}}
     \STATE {$D_{i} \gets {\sf F}(M_{i-1},{\cal D}_{i-1},{\sf C})$}~~\COMMENT{manipulated version of $D_{0}$}
\ENDIF
\IF{$i<\alpha$}
\STATE $M_{i} \gets {\sf MLA}(D_{i})$ \COMMENT{$D_1,\ldots,D_{\alpha-1},M_1,\ldots,M_{\alpha-1}$ are not used when {\sf ST}=={\tt parallel-adaptation}}
\ENDIF
\ENDFOR
\RETURN $D_\alpha$
\end{algorithmic}
}
\end{algorithm}

Algorithm \ref{resilience-master-attacker} describes the adaptive attack. 
As highlighted in Figure \ref{fig:attack-adaptation-methodologies}, we consider three basic adaptation strategies.
\begin{itemize}
\item {\sf ST} == {\tt parallel-adaptation}:
The attacker sets the manipulated $D_{i}={\sf F}(M_{0},D_{0},{\sf C})$, where $i=1,\ldots,\alpha$, and ${\sf F}$ is a randomized manipulation algorithm,
meaning that $D_{i} =D_{j}$ for $i\neq j$ is unlikely.

\item {\sf ST} == {\tt sequential-adaptation}:
The attacker sets the manipulated $D_{i}={\sf F}(M_{i-1},D_{i-1},{\sf C})$ for $i=1,\ldots,\alpha$,
where detection schemes $M_{1},\ldots,M_\alpha$ are respectively learned from $D_1,\ldots,D_\alpha$ using the defender's machine learning algorithm ${\sf MLA}$
(also known to the attacker).

\item {\sf ST} == {\tt full-adaptation}:
The attacker sets the manipulated
$D_{i}={\sf F}(M_{i-1},{\sf PP}(D_0,\ldots,D_{i-1}),{\sf C})$ for $i=1,2,\ldots$,
where ${\sf PP}(\cdot,\ldots)$ is a preprocessing algorithm for ``aggregating"  sets of feature vectors $D_0,D_1,\ldots$
into a single set of feature vectors, ${\sf F}$ is a manipulation algorithm,
$M_1,\ldots,M_\alpha$ are learned respectively from $D_1,\ldots,D_\alpha$ by the attacker. 
\end{itemize}

Algorithm \ref{preproc} is a concrete preprocessing algorithm.
Its basic idea is the following:
since each malicious website corresponds to $m$ malicious feature vectors that respectively belong to
$D_0,\ldots,D_{m-1}$,
the preprocessing algorithm randomly picks one of the $m$  malicious feature vectors to represent the malicious website in ${\cal D}$.
It is worth mentioning that one can derive some hybrid attack strategies from the above three basic strategies.
We also note that the attack strategies and the manipulation constraints are independent of the detection schemes,
but the manipulation algorithms would be specific to the detection schemes.

\begin{algorithm}[hbtp]
{\small
\caption{Preprocessing ${\sf PP}(D_0,\ldots,D_{m-1})$}
\label{preproc}
INPUT: $m$ sets of feature vectors $D_0,\ldots,D_{m-1}$ where the $z$th malicious website corresponds to $D_0.malicious[z],\ldots,D_{m-1}.malicious[z]$\\
OUTPUT: ${\cal D}$
\begin{algorithmic}[1]
\STATE ${\cal D}\gets \emptyset$
\STATE{{\tt size} $\gets$ {\sf sizeof}($D_{0}.malicious$)} 
\FOR{$z=1$  $\TO$ {\tt size}}
\STATE{${\cal D}[z] \stackrel{R}{\gets} \{D_{0}.malicious[z],\ldots, D_{m-1}.malicious[z]\}$}
\STATE{${\cal D}\gets{\cal D}\cup D_0.benign$}
\ENDFOR
\RETURN ${\cal D}$
\end{algorithmic}
}
\end{algorithm}

\noindent{\bf Manipulation Constraints.}
For a feature $X$ whose value is to be manipulated, the attacker needs to compute $X.escape\_interval$,
which is a subset of feature $X$'s domain $domain(X)$ and can possibly cause the malicious feature vector to evade detection.
Feature $X$'s manipulated value is randomly chosen from its $escapte\_interval$, which is calculated using
Algorithm \ref{alg:resilience-constraints}, while taking as input $X$'s
domain constraints and semantics constraints.

\begin{algorithm}[hbt]
{\small
\caption{$X$'s escape\_interval ${\sf Escape}(X,M,{\sf C})$}
\label{alg:resilience-constraints}
INPUT: $X$ is feature for manipulation,
$M$ is detection scheme,
${\sf C}$ represents constraints \\
OUTPUT: $X$'s $escape\_interval$
\begin{algorithmic}[1]
\STATE{$domain\_constraint \gets {\sf C}.domain\_map(X)$}
\STATE{$semantics\_constraint \gets {\sf C}.semantics\_map(X)$}~~\COMMENT{$\emptyset$ if $X$ cannot be manipulate due to semantics constraints}
\STATE{$escape\_interval \gets domain\_constraint\cap semantics\_constraint$}
\RETURN $escape\_interval$
\end{algorithmic}
}
\end{algorithm}

Algorithm \ref{alg:resilience-constraints} is called because the manipulation algorithm needs to
compute the interval from which a feature's manipulated value should be taken.
Specifically, the constraints are the following.
\begin{itemize}
\item {\bf Domain constraints}:
Each feature has its own domain of possible values.
This means that the new value of a feature after manipulation must fall into the domain of the feature.
Let ${\sf C}.domain\_map$ be a table of $(key, value)$ pairs, where $key$ is feature name and $value$ is the feature's domain constraint.
Let ${\sf C}.domain\_map(X)$ return feature $X$'s domain as defined in ${\sf C}.domain\_map$.

\item {\bf Semantics constraints}:
The manipulation of feature values should have no side-effect to the attack,
or at least cannot invalidate the attacks.
For example, if a malicious website needs to use script to launch the drive-by-download attack,
the feature indicating the number of scripts 
cannot be manipulated to 0.
Let ${\sf C}.semantics\_map$ be a table of $(key, value)$ pairs, where $key$ is feature name and $value$ is the feature's {\em semantics constraints}.
Let ${\sf C}.semantics\_map(X)$ return feature $X$'s semantics constraints as specified in ${\sf C}.attack\_map$.

\ignore{%!}
\item {\bf Correlation constraints}:
Some features may be correlated to each other. This means that these features' values should not be manipulated independently of each other;
otherwise, adaptive attacks can be defeated by simply examining the violation of correlations.
Correlation constraints can be automatically derived from data on demand (as done in our experiments), or alternatively given as input.
Let ${\sf C}.group$ be a table of $(key, value)$ pairs, where $key$ is feature name and $value$ records the feature's correlated features.
Let ${\sf C}.group(X)$ return the set of features belonging to ${\sf C}.group$,
namely the features that are correlated to $X$.
}%!
\end{itemize}

In general, constraints might have to be manually identified based on feature definitions and domain knowledge.

\begin{figure}[!hbtp]
\centering
\includegraphics[width=0.45\textwidth]{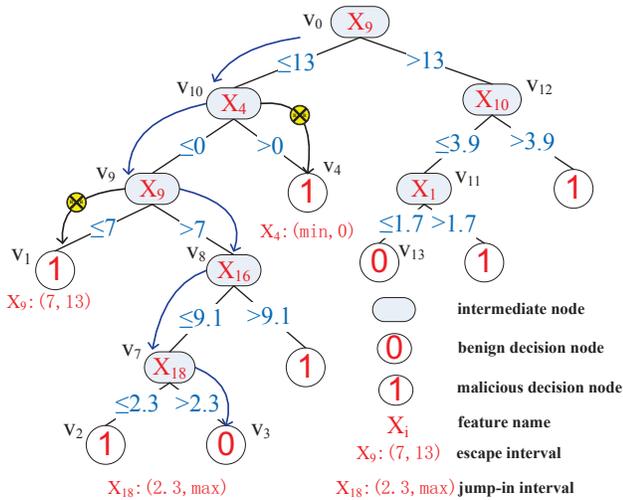}
\caption{Example J48 classifier and feature manipulation.
For inner node $v_{10}$ on the $benign\_path$ ending at $benign\_leaf$ $v_3$, we have
$v_{10}.feature=``X_4"$ and $v_{10}.feature.value=X_4.value$.
\label{fig:simple-attack-scenario}}
\vspace{-0.1in}
\end{figure}

\noindent{\bf Manipulation Algorithms.}
As mentioned in Section \ref{sec:data-collection}, we adopt the 
J48 classifier detection scheme,
where a J48 classifier is trained by concatenating the application- and network-layer features corresponding to the same URL \cite{codaspy13:li}.
We present two manipulation algorithms, called ${\sf F}_1$ and ${\sf F}_2$,
which exploit the defender's J48 classifier to guide the manipulation of features.
Both algorithms neither manipulate the benign feature vectors
(which are not controlled by the attacker), nor manipulate the malicious feature vectors
that are already classified as benign by the defender's detection scheme (i.e., false-negative).
Both algorithms may fail,
while brute-forcing may fail as well because of the manipulation constraints.

Since the manipulation algorithms are inevitably complicated, in the following we will present their basic ideas and sketched algorithms.
The notations used in the algorithms are: 
for node $v$ in the classifier, $v.feature$ is the feature associated to node $v$,
and $v.value$ is $v.feature$'s ``branching" value as specified by the classifier (a binary tree with all features numericalized).

\begin{algorithm}[!hbtp]
{\small
\caption{Manipulation algorithm ${\sf F}_1(M,D,{\sf C})$}
\label{resilience-alg1}
INPUT: J48 classifier $M$, feature vector set $D$(malicious $\cup$ benign), manipulation constraints $C$ \\
OUTPUT: manipulated feature vectors
\begin{algorithmic}[1]
\FORALL{feature vector $\in$ $D$.malicious}
    \STATE $v$ be the root node of $M$
    \STATE maintain an interval for every feature in feature vector.
    \WHILE{ $v$ is not benign leaf }
        \IF { $v$ is an inner node}
            \STATE {$v \gets v.Child$ based on decision tree rule}
            \STATE{update corresponding feature interval}
        \ELSIF {$v$ is a malicious leaf}
            \STATE{compute the corresponding feature's {\em escape\_interval} by calling Algorithm \ref{alg:resilience-constraints}}
            \STATE pick a value $z$ in the {\em escape\_interval} uniformly at random
            \STATE {set the corresponding feature's value to $z$}
            \STATE {$v \gets v.sibling $}
        \ENDIF
    \ENDWHILE
\ENDFOR
\RETURN manipulated feature vectors $D$
\end{algorithmic}
}
\end{algorithm}

Manipulation Algorithm ${\sf F}_1$ is described as Algorithm \ref{resilience-alg1}. 
The basic idea underlying this manipulation algorithm
is the following: for every malicious feature vector in $D$, there is a unique path (in the J48 classifier $M$)
that leads to a {\em malicious leaf}, which indicates that the feature vector is malicious.
We call the path leading to malicious leaf a {\em malicious path}, and the path leading to a {\em benign leaf}
(which indicates a feature vector as benign) a {\em benign path}.

By examining the path from the malicious leaf to the root, say $malicious\_leaf\rightarrow v_2\rightarrow \ldots \rightarrow root$,
and identifying the first inner node, namely $v_2$, the algorithm attempts to manipulate ${\sf fv}.(v_2.feature).value$ so that the classification
can lead to $malicious\_leaf$'s sibling, say $v_{2,another\_child}$, which is guaranteed to exist (otherwise, $v_2$ cannot be an inner node).
Note that there must be a sub-path rooted at $v_{2,another\_child}$ that leads to a $benign\_leaf$ (otherwise, $v_2$ cannot be an inner node as well),
and that manipulation of values of the features corresponding to the nodes on the 
sub-tree rooted at $v_{2,another\_child}$ will preserve the postfix $v_2\rightarrow \ldots \rightarrow root$.

To help understand the manipulation algorithm,
let us look at one example.
At a high-level, the attacker runs ${\sf AA}(``J48", M_0,D_0,{\sf ST}, {\sf C},{\sf F}_1,\alpha=1)$
and therefore ${\sf F}_1(M_0,D_0,{\sf C})$ to manipulate the feature vectors,
where {\sf ST} can be any of the three strategies because
they cause no difference when $\alpha=1$ (see Figure \ref{fig:attack-adaptation-methodologies}).
Consider the example J48 classifier $M$ in Figure \ref{fig:simple-attack-scenario}, where features and their values are for illustration purpose,
and the leaves are decision nodes with class 0 indicating {\em benign leaves} and 1
indicating {\em malicious leaves}.
A website with feature vector
$$(X_{4}=-1,X_{9}=5,X_{16}=5,X_{18}=5)$$
is classified as malicious because it leads to decision path
$${v_{0} \xrightarrow{X_9\le13} v_{10} \xrightarrow{X_4\le0} v_{9} \xrightarrow{X_9\le7} v_{1}},$$
which ends at malicious leaf $v_1$.
The manipulation algorithm first identifies malicious leaf $v_1$'s parent node $v_9$,
and manipulates $X_9$'s value to fit into $v_1$'s sibling ($v_8$).
Note that $X_9$'s {\em escape\_interval} is as:
\begin{eqnarray*}
([\min\nolimits_{9}, \max\nolimits_{9}]\setminus [\min\nolimits_{9}, 7])\cap [\min\nolimits_{9}, 13] =(7, 13],
\end{eqnarray*}
where $Domain(X_9)=[\min_9,\max_9]$,
$[\min_9,7]$ corresponds to node $v_9$ on the path, and $[\min_0,13]$ corresponds to node $v_0$ on the path.
The algorithm manipulates $X_9$'s value to be a random element from $X_9$'s $escapte\_interval$, say $8\in(7,13]$,
which causes the manipulated feature vector to evade detection because of decision path:
$$v_{0} \xrightarrow{X_9\le13} v_{10} \xrightarrow{X_4\le0} v_{9} \xrightarrow {X_9 >7}
v_{8} \xrightarrow{X_{16}\le9.1} v_{7} \xrightarrow{X_{18}>2.3} v_{3}$$
and ends at benign leaf $v_3$.

Manipulation Algorithm ${\sf F}_2$ is described as Algorithm \ref{resilience-alg2}. 
The basic idea underlying this manipulation algorithm is to first extract all benign paths.
For each feature vector ${\sf fv}\in D.malicious$,
${\sf F}_2$ keeps track of the mismatches between ${\sf fv}$ and all benign paths. 
The algorithm attempts to manipulate as few ``mismatched" features as possible to evade $M$.

\begin{algorithm}[!hbtp]
{\small
\caption{Manipulation algorithm ${\sf F}_2(M,D,{\sf C})$}
\label{resilience-alg2}
INPUT: J48 classifier $M$, feature vector set $D$(malicious $\cup$ benign), manipulation constraints $C$ \\
OUTPUT: manipulated feature vectors
\begin{algorithmic}[1]
\STATE create interval vector for features along every benign path and store in $\sf Paths$.
\FORALL{feature vector $\in$ $D$.malicious}
    \FORALL{$Path \in  {\sf Paths}$}
        \STATE{compare feature vector to $Path$ and record the mismatch feature number}
    \ENDFOR
    \STATE{sort $\sf Paths$ in ascending order of mismatch feature number}

    \FORALL{$Path \in {\sf Paths}$ {\bf until} successfully manipulated}
        \FORALL{mismatch feature $\in$ $\sf Paths$}
            \STATE{get escape interval by calling Algorithm \ref{alg:resilience-constraints}}
            \STATE pick a value $n$ in escape interval at random
            \STATE {set feature vector's corresponding feature value to $n$}
        \ENDFOR
    \ENDFOR
\ENDFOR
\RETURN manipulated feature vectors $D$
\end{algorithmic}
}
\end{algorithm}

To help understand this manipulation algorithm, let us look at another example.
Consider feature vector
$$(X_4=.3,X_9=5.3,X_{16}=7.9,X_{18}=2.1,X_{10}=3,X_{1}=2.3),$$
which is classified as malicious because of path
$${v_{0} \xrightarrow{X_9\le13} v_{10} \xrightarrow{X_4>0} v_{4}}.$$
To evade detection, the attacker can compare the feature vector to the matrix of two benign paths.
For the benign path $v_3 \rightarrow v_7 \rightarrow v_8 \rightarrow v_9 \rightarrow v_{10} \rightarrow v_0$,
the feature vector has three mismatches, namely features $X_4,X_9,X_{18}$.
For the benign path $v_{13} \rightarrow v_{11} \rightarrow v_{12} \rightarrow v_{0}$, the feature vector has two mismatches, namely  $X_9$ and $X_1$.
The algorithm first processes the benign path ending at node $v_{13}$.
The algorithm will try to manipulate $X_9$ and $X_1$'s values to reach the benign leaf.
Suppose on the other hand, that $X_{10}$ cannot be manipulated without violating the constraints.
The algorithm stops with this benign path and considers the benign path end at node $v_3$.
If the algorithm fails with this benign path again, the algorithm will not manipulate the feature vector
and leave it to be classified as malicious. %! by defender's J48 classifier $M$.

\ignore{ %!
\begin{algorithm}[!hbtp]
{\small
\caption{Maintaining constraints ${\sf MR}(M, {\sf C}, S)$}
\label{maintain-remains}
INPUT: J48 classifier $M$, manipulation constraints ${\sf C}$, $S=\{(feature,value,interval,manipulated)\}$ \\
OUTPUT: ${\tt true}$ or ${\tt false}$
\begin{algorithmic}[1]
    \STATE {$S^* \gets \{s\in S:s.manipulated=={\tt true}\}$}
    \FORALL{$(feature,value,interval,{\tt true}) \in S$}
        \FORALL{$X \in {\sf C}.group(feature) \setminus S^*.features$}
            \STATE{$\{X_1,\ldots,X_j\}\gets {\sf C}.group(feature)\cap S^*.features$, whose values are respectively $x_1,\ldots,x_j$ w.r.t. $S^*$}
            \STATE{$escape\_interval \gets {\sf Escape}(feature, M, {\sf C}, (X_1=x_1,\ldots,X_j=x_j))$}
            %\STATE{$escape\_interval \gets  {\sf Constraint}(x, M, {\sf C}, S)$}
            \IF{$escape\_interval==\emptyset$}
                \RETURN{${\tt false}$}
            \ELSE
                \STATE{$X.interval \gets escape\_interval$}
                \STATE{$X.value \stackrel{R}{\gets} X.interval$}
                %\STATE{$X.manipulated \gets {\tt true}$}
                \STATE{$S^* \gets S^* \cup \{(X,X.value,X.interval,{\tt true})\}$}
            \ENDIF
        \ENDFOR
    \ENDFOR
    \RETURN{${\tt true}$}
\end{algorithmic}
}
\end{algorithm}
}%!

\subsection{Counter-Evasion Algorithms}

We have showed that adaptive attacks can ruin the defender's (non-proactive) detection schemes.
Now we investigate counter-measure against adaptive attacks.
\begin{algorithm}[!hbtp]
{\small
\caption{Proactive detection ${\sf PD}({\sf MLA}, M_0, {D}_0^\dag,D_\alpha, {\sf ST}_D,{\sf C},{\sf F}_D,\gamma)$}
\label{protect-detectin-algorithm}
INPUT: $M_0$ is learned from $D'_0$ using {\sf MLA},
$D_0^\dag=D_0^\dag.benign\cup D_0^\dag.malicious$,
${D}_\alpha$ ($\alpha$ unknown to defender) is set of feature vectors for classification (where the malicious websites may have been manipulated by the
attacker),
${\sf ST}_D$ is defender's adaptation strategy,
${\sf F}_D$ is defender's manipulation algorithm,
{\sf C} is set of constraints,
$\gamma$ is defender's number of adaptations rounds  \\
OUTPUT: malicious  vectors ${\sf fv}\in D_\alpha$
\begin{algorithmic}[1]
\STATE{$M_1^\dag,\ldots,M_\gamma^\dag\gets {\sf PT}({\sf MLA}, M_0, D_0^\dag, {\sf ST}_D, {\sf C},{\sf F}_D,\gamma)$} ~~\COMMENT{see Algorithm \ref{fig:pda-master}}
\STATE{${\sf malicious} \gets \emptyset$}
\FORALL{${\sf fv} \in D_\alpha$}
    \IF{($M_0({\sf fv})$ says {\sf fv} is malicious) OR (majority of $M_0({\sf fv}),M_1^\dag({\sf fv}),\ldots,M_\gamma^\dag({\sf fv})$ say {\sf fv} is malicious)}
        \STATE{${\sf malicious} \gets {\sf malicious} \cup \{{\sf fv}\}$}
    \ENDIF
\ENDFOR
\RETURN ${\sf malicious}$
\end{algorithmic}
}
\end{algorithm}
The counter-measure is based on the idea of {\em proactive training \& detection}.
Algorithm \ref{protect-detectin-algorithm} describes the {\em proactive detection} algorithm.
The basic idea of this algorithm is to call the {\em proactive training} algorithm to generate a
set of proactively trained detection schemes, denoted by $M_1^\dag,\ldots,M_\gamma^\dag$.
These detection schemes are derived from, among other things, $M_0$, which is learned from $D'_0$ using {\sf MLA}.
It is important to note that $D'_0 =D'_0.benign\cup D'_0.malicious$,
where $D'_0.benign$ is a set of websites that are known to be benign (ground truth) and cannot be manipulated by the attacker,
$D'_0.malicious$ is a set of websites that are known to be malicious (ground truth).
A website is classified as malicious if the non-proactive detection scheme $M_0$ classifies it as malicious, or
at least $\lfloor (\gamma+1)/2 \rfloor + 1$ of the proactively trained detection schemes classify it as malicious.
This is meditated to accommodate that the defender does not know {\em a priori} whether the attacker is adaptive or not,
When the attacker is not adaptive, $M_0$  can effectively deal with $D_0$.

Algorithm \ref{fig:pda-master} describes the proactive training algorithm.
This algorithm is similar to the adaptive attack algorithm {\sf AA}
because it also consider three kinds of adaptation strategies.
Specifically, this algorithm aims to derive detection schemes $M_1^\dag,\ldots,M_\gamma^\dag$
from the starting-point detection scheme $M_0$.

\begin{algorithm}[!hbtp]
{\small
\caption{Proactive training ${\sf PT}({\sf MLA}, M_0, D_0^\dag, {\sf ST}_D, {\sf C},{\sf F}_D,\gamma)$}
\label{fig:pda-master}
INPUT: same as in {\bf Algorithm \ref{protect-detectin-algorithm}} \\
OUTPUT: $M_1^\dag, \ldots, M_{\gamma}^\dag$
\begin{algorithmic}[1]
\STATE{$M_0^\dag\gets M_0$}~~\COMMENT{for simplifying notations}
\STATE initialize $D_1^\dag,\ldots,D_{\gamma}^\dag$ and $M_1^\dag,\ldots,M_{\gamma}^\dag$ respectively as empty sets and empty classifiers
\FOR{$i$=1 $\TO$ $\gamma$}
\IF{${\sf ST}_D == {\tt parallel\text{-}adaptation}$}
  \STATE{$D_{i}^\dag.malicious \gets {\sf F}_D(M_{0}^\dag,D_{0}^\dag.malicious,{\sf C})$}
\ELSIF{${\sf ST}_D == {\tt sequential\text{-}adaptation}$}
   \STATE {$D_{i}^\dag.malicious \gets {\sf F}_D(M_{i-1}^\dag,D_{i-1}^\dag.malicious,{\sf C})$}
 \ELSIF{${\sf ST}_D == {\tt full\text{-}adaptation}$}
     \STATE {${\cal D}_{i-1}^\dag.malicious \gets {\sf PP}(D_0^\dag,\ldots,D_{i-2}^\dag)$}
     \STATE {$D_{i}^\dag.malicious \gets {\sf F}_D(M_{i-1}^\dag,{\cal D}_{i-1}^\dag,{\sf C})$}
\ENDIF
\STATE $D_i^\dag.benign\gets D_0^\dag.benign$
\STATE {$M_{i}^\dag \gets {\sf MLA}(D_i^\dag)$}
\ENDFOR
\RETURN $M_1^\dag, \ldots, M_\gamma^\dag$
\end{algorithmic}
}
\end{algorithm}

\begin{algorithm}[!hbtp]
{\small
\caption{Proactive defense vs. adaptive attack evaluation \newline
${\sf Eva}({\sf MLA}, M_0, D_0^\dag,{D}_0, {\sf ST}_A, {\sf F}_A,{\sf ST}_D,{\sf F}_D,{\sf C},\alpha,\gamma)$}
\label{protect-defense-vs-adaptive-attack-algorithm}
INPUT: detection scheme $M_0$ (learned from $D'_0$, as in Algorithm \ref{fig:pda-master}),
$D_0^\dag$ is set of feature vectors for defender's proactive training,
${D}_0=D_0.malicious\cup D_0.benign$,
${\sf ST}_A$ (${\sf ST}_D$) is attacker's (defender's) adaptation strategy,
${\sf F}_A$ (${\sf F}_D$) is attacker's (defender's) manipulation algorithm,
{\sf C} is the constraints,
$\alpha$ ($\gamma$) is the number of attacker's (defender's) adaptation rounds  \\
OUTPUT: ACC, FN, TP and FP
\begin{algorithmic}[1]
\IF{$\alpha>0$}
  \STATE{$D_\alpha\gets {\sf AA}({\sf MLA}, M_0, D_0, {\sf ST}_A, {\sf C}, {\sf F}_A,\alpha)$}  \newline ~~\COMMENT{call Algorithm \ref{resilience-master-attacker}}
\ENDIF
\STATE{$M_1^\dag,\ldots,M_\gamma^\dag\gets {\sf PT}({\sf MLA}, M_0, D_0^\dag, {\sf ST}_D,{\sf C}, {\sf F}_D,\gamma)$} \newline ~~\COMMENT{call Algorithm \ref{fig:pda-master}}
\STATE{${\sf malicious}\gets {\sf PD}({\sf MLA}, M_0, {D}_0^\dag,D_\alpha, {\sf ST}_D,{\sf C},{\sf F}_D,\gamma)$} \newline ~~\COMMENT{call Algorithm \ref{protect-detectin-algorithm}}
\STATE{${\sf benign}\gets D_\alpha\setminus {\sf malicious}$}
\STATE{calculate ACC, FN, TP and FP w.r.t. $D_0$}
\RETURN ACC, FN, TP and FP
\end{algorithmic}
}
\end{algorithm}
Algorithm \ref{protect-defense-vs-adaptive-attack-algorithm} describes the algorithm for evaluating the effectiveness of the counter-measure against the adaptive attacks.
Essentially, the evaluation algorithm calls the defender's protection detection to generate a set of proactively trained detection schemes,
and calls the attacker's adaptive attack algorithm to manipulate the malicious websites (i.e., selecting some of their features and manipulating their values
to evade a given detection scheme).
By varying the adaptation strategies and the parameters, we can evaluate the
effectiveness of proactive training \& detection against the adaptive attacks.
The parameter space of the evaluation algorithm includes at least 108 scenarios: the basic adaptation strategy space ${\sf ST}_A\times {\sf ST}_D$
is $3\times 3$ (i.e., not counting any hybrids
of ${\tt parallel\text{-}adaptation}$, ${\tt sequential\text{-}adaptation}$ and ${\tt full\text{-}adapatation}$),
the manipulation algorithm space ${\sf F}_A\times {\sf F}_B$ is $2 \times 2$, and the adaptation round parameter space is at least $3$ ($\alpha>,=,<\gamma$).

\section{Evaluating Effectiveness of the Counter-Evasion Algorithms}
\label{sec:evaluating-framework}

We use the standard metrics, including false-negative and false-positive rates \cite{codaspy13:li}, to 
evaluate the effectiveness of counter-evasion algorithms. 

\subsection{Data Description}

The dataset used in this paper consists of a 40-day URLs. 
Malicious URLs are downloaded from blacklists: \url{compuweb.com/url-domain-bl.txt}, \url{malware.com.br}, \url{malwaredomainlist.com}, \url{zeustracker.abuse.ch} and \url{spyeyetracker.abuse.ch} and further confirmed with the high-interaction client honeypot Capture-HPC\cite{citeulike:5765320}. Benign URLs are obtained from \url{alexa.com}, which lists the top 10,000 websites that are supposed to be well protected. The test of blacklist URLs using high-interaction client honeypot confirmed our observation that some or many blacklist URLs are not accessible any more and thus should not be counted as malicious URLs. 

\ignore{
The dataset for our case study was collected by an automatical script, which will routinely obtained benign and malicious URL list for 40 days period of time, which happened in consecutive days from 02/01/2012 to 03/11/2012. The malicious URLs list are downloaded from blacklists and are confirmed as malicious by the high-interaction client honeypot Capture-HPC version 3.0  \cite{citeulike:5765320}. The malicious URL list are from the following blacklists:
\url{compuweb.com/url-domain-bl.txt}, \url{malware.com.br}, \url{malwaredomainlist.com}, \url{zeustracker.abuse.ch} and \url{spyeyetracker.abuse.ch}. Although the blacklists are updated frequently, some malicious websites have very short life time, which are not accessible or malicious by the time when we tried to access it. Although high-interaction client honeypot has false positive, consider the fact that these malicious URLs are from black list and our experiment is conducted at the same day blacklist URLs are released, so it's safe to say all the malicious URLs of our dataset are actually malicious.
The test of blacklist URLs using high-interaction client honeypot confirmed our observation that some or many blacklist URLs are not accessible any more and thus should not be counted as malicious URLs. The benign URLs are obtained from \url{alexa.com}, which lists the top 10,000 websites that are supposed to be well protected. By eliminating non-accessible benign websites, the average number of benign websites in 40 days are 9501.
}%EOI

The daily average number of malicious websites listed in blacklists is 6763 
 and the daily average number of malicious websites after being verified by Capture-HPC is 838. The total number of distinct malicious websites we found in 40 days are 17091. The daily average number of benign websites is 10,000. By eliminating non-accessible benign websites, the daily average number of benign websites is 9501 in 40 days.
According to \cite{url:oakland:11} and our experiment results, it can achieve best detection rate, when the rate between the number of benign websites and malicious websites are 4:1. So we choose all malicious websites and 4 times of benign websites as our training and testing data.

\ignore{%!
\begin{table}[!hbtp]
\caption{Data summary}
\label{table:data-description}
\centering
{\scriptsize
\begin{tabular}{|c|c|c|c|}
\hline
Website  & Daily Ave.  \# of        & Daily Ave. \# of   & Total \# of Distict \tabularnewline
Category &  Websites  &  Verified Websites  & Websites \tabularnewline
\hline
Benign Websites& 10,000 & 9,501  & 10,000  \tabularnewline
\hline
Malicious Websites& 6,763   & 838     & 17,091  \tabularnewline
\hline
\end{tabular}
}

\end{table}
} %!

\subsection{Effectiveness of the Evasion Attacks}
\label{sec:power-of-adaptive-attack}

%end of ignore.

Table \ref{table:attack-power} summarizes the results of adaptive attack ${\sf AA}(``J48",M_0,$ $D_0,{\sf ST},{\sf C},{\sf F},\alpha=1)$
based on the 40-day dataset mentioned above. 
The experiment can be more succinctly represented as $M_0(D_1)$, meaning that the defender is static (or non-proactive) and
the attacker is adaptive with $\alpha=1$, where $D_1$ is the manipulated version of $D_0$.
Note that in the case of $\alpha=1$, the three adaptation strategies lead to the same $D_1$ as shown in Figure \ref{fig:attack-adaptation-methodologies}.
From Table \ref{table:attack-power}, we find that both manipulation algorithms can effectively evade detection
by manipulating on average 4.31-7.23 features while achieving false-negative rate 87.6\%-94.7\% for ${\sf F}_1$,
and by manipulating on average 4.01-6.19 features while achieving false-negative rate 89.1\%-95.3\% for ${\sf F}_2$
\begin{table}[!htbp]
\caption{Experiment results with $M_0(D_1)$
in terms of average false-negative rate (FN),
average number of manipulated features (\#MF), average percentage of failed attempts (FA).
\label{table:attack-power}}
\centering
{\scriptsize
\begin{tabular}{|l|r|r|r|r|r|r|}
\hline
    & \multicolumn{3}{c|}{${\sf F_1}$} & \multicolumn{3}{c|}{${\sf F_2}$} \\
\hline
  Detection Scheme &  FN & \#MF &FA &  FN & \#MF & FA \\
\hline
J48 Decision Tree&   87.6\% & 7.23 & 12.6\%  & 89.1\% & 6.19& 11.0\%\\
\hline
\end{tabular}
}
\vspace{-0.1in}
\end{table}

Having observed the phenomenon that manipulation of some features' values can essentially make the detection schemes useless,
it would be natural to ask {\bf{\em which features are often manipulated for evasion?}} 
To look into the question, we notice that many features are manipulated over the 40 days, but only a few are manipulated often.

%Figure \ref{fig:appmanipulate} summarizes the application-layer features that are manipulated to bypass classifier $M_{0}$.

\ignore{

For application-layer alone, ${\sf F}_1$ most often (i.e., $>150$ times each day for over the 40 days) manipulates the following five application-layer features:
URL length ({\tt URL\_length}), number of scripts contained in website content ({\tt \#Script}),
webpage length ({\tt Content\_length}), number of URLs embedded into the website contents ({\tt \#Embedded\_URL}),
and number of Iframes contained in the webpage content ({\tt \#Iframe}).
In contrast, ${\sf F}_2$ most often (i.e., $>150$ times) manipulates the following three application-layer features:
number of special characters contained in URL ({\tt \#Special\_character}),
number of long strings ({\tt \#Long\_strings}) and
webpage content length
({\tt Content\_length}).
That is, {\tt Content\_length} is the only feature that is most often manipulated by both algorithms.

For network-layer alone, ${\sf F}_1$ most often (i.e., $>150$ times) manipulates the following three features:
number of remote IP addresses ({\tt \#Dist\_remote\_IP}),
duration time ({\tt Duration}), and number of application packets ({\tt \#Local\_app\_packet}).
Whereas, ${\sf F}_2$ most often (i.e., $>150$ times) manipulates the distinct number of TCP ports used by the remote servers ({\tt \#Dist\_remote\_TCP\_port}).
In other words, no single feature is often manipulated by both algorithm.

}

${\sf F}_1$ most often (i.e., $>150$ times each day for over the 40 days) manipulates
three application-layer features ---  {\tt URL\_length}, {\tt Content\_length},
{\tt \#Embedded\_URLs} ---
and two network-layer features --- {\tt Duration} and  
{\tt \#Local\_app\_packet}.
On the other hand, ${\sf F}_2$ most often (i.e., $>150$ times) manipulates
two application-layer features ---  {\tt \#Special\_characters}
and {\tt Content\_length} --- and one network-layer feature --- {\tt Duration}.

The above discrepancy between the frequencies that features are manipulated
can be attributed to the design of the manipulation algorithms.
Specifically, ${\sf F}_1$ seeks to manipulate features that are associated to nodes that are close to the leaves.
In contrast, ${\sf F}_2$ emphasizes on the mismatches between a malicious feature vector and an entire benign path,
which represents a kind of global search and also explains why ${\sf F}_2$ manipulates fewer features.

We also want to know {\bf{\em why these features have such high security/evasion significance?}} 
The issue is important because identifying the ``important" features could lead to deeper insights. 
We compare the manipulated features to the features that would be selected by a feature selection 
algorithm for the purpose of training classifiers. To be specific, we use the {\sf InfoGain} feature selection algorithm
because it ranks the contributions of individual features \cite{codaspy13:li}.
We find that among the manipulated features, {\tt URL\_length} is the only feature among the five {\sf InfoGain}-selected application-layer features,
and {\tt \#Dist\_remote\_TCP\_port} is the only feature among the four {\sf InfoGain}-selected network-layer features. 
This suggests that the feature selection algorithm does not necessarily offer good insights into
the importance of features from a security perspective.

The standard feature-selection algorithms are almost useless to find indicative features from the perspective 
of evading classifiers. There is still gap between our results and the ``optimal" solutions based on security semantics; 
it's an open problem to bridge the gap, because classifiers are ``black-box" that don't really accommodate 
``security semantics" of features.

%end of ignore

\subsection{Effectiveness of the Counter-Evasion Algorithms}

Table \ref{table:cross-and-aggregation} summarizes the effectiveness
of proactive defense against adaptive attacks.
We make the following observations.
First, if the defender is proactive (i.e., $\gamma>0$) but the attacker is non-adaptive (i.e., $\alpha=0$),
the false-negative rate drops from 0.79\% in the baseline case
to some number belonging to interval $[0.23\%,0.56\%]$.
The price is: the detection accuracy drops from 99.68\% in the baseline case to some number belonging to interval $[99.23\%,99.68\%]$
the false-positive rate increases from 0.14\% in the baseline case to some number belonging to interval $[0.20\%,0.93\%]$,
The above observations suggest: {\bf the defender can always use proactive detection without worrying about
side-effects (e.g., when the attacker is not adaptive)}. This is because the proactive detection algorithm {\sf PD}
uses $M_0(D_0)$ as the first line of detection.

\begin{table*}[thb]
\centering
\caption{Cross-layer proactive detection with ${\sf ST}_A={\sf ST}_D$.
For baseline case $M_0(D_0)$, ACC = 99.68\%, true-positive rate TP =99.21\%,
false-negative rate FN=0.79\%, and false-positive rate FP=0.14\%.
\label{table:cross-and-aggregation}}
{\scriptsize
\begin{tabular}{|l|l|rrrr|rrrr|rrrr|}
\hline
\multirow{2}{*}{Strategy} & \multirow{2}{*}{Manipulation algorithm}  & \multicolumn{4}{c|}{ $M_{0\text{-}8}(D_0)$ }  & 
\multicolumn{4}{c|}{$M_{0\text{-}8}(D_1)$} &\multicolumn{4}{c|}{$M_{0\text{-}8}(D_9)$ }\\
\cline{3-14}
            &  &ACC & TP & FN & FP          &ACC & TP & FN & FP        &ACC  & TP & FN & FP \\
\hline
\multirow{3}{*}{${\sf ST}_A={\sf ST}_D$} & ${\sf F}_D={\sf F}_1$ vs. ${\sf F}_A={\sf F}_1$   & 99.59& 99.71  & 0.29  
& 0.39 & 95.58 & 92.03 & 7.97 & 3.62   & 95.39 & 92.00  & 8.00   & 3.83 \\ \cline{2-14}
\multirow{4}{*}{$={\tt parallel}$}  & ${\sf F}_D={\sf F}_1$ vs. ${\sf F}_A={\sf F}_2$  															& 99.27& 99.77  & 0.23  & 0.77 & 78.51 & 25.50 & 74.50 & 9.88  & 78.11 & 32.18  & 67.82  & 11.48    \\ \cline{2-14}
   & ${\sf F}_D={\sf F}_2$ vs. ${\sf F}_A={\sf F}_1$  																		
   & 99.16& 99.76  & 0.24  & 0.93 & 76.33 & 19.32 & 80.68 & 11.17 & 78.96 & 39.77  & 60.23  & 12.14          \\ \cline{2-14}
   & ${\sf F}_D={\sf F}_2$ vs. ${\sf F}_A={\sf F}_2$  & 99.59& 99.62  & 0.38  & 0.39 & 93.66 & 90.25 & 9.75 & 5.59   & 96.17 & 92.77  & 7.23   & 3.08          \\ \hline
\multirow{3}{*}{${\sf ST}_A={\sf ST}_D$} & ${\sf F}_D={\sf F}_1$ vs. ${\sf F}_A={\sf F}_1$ & 99.52& 99.69  & 0.31  & 
0.45 & 93.44 & 77.48 & 22.52 & 3.05  & 92.04 & 59.33  & 30.67  & 2.99 \\ \cline{2-14}
\multirow{4}{*}{$={\tt sequential}$}  & ${\sf F}_D={\sf F}_1$ vs. ${\sf F}_A={\sf F}_2$  														& 99.23& 99.70  & 0.30  & 0.82 & 74.24 & 20.88 & 79.22 & 14.06 & 79.43 & 30.03  & 69.97  & 9.38          \\ \cline{2-14}
   & ${\sf F}_D={\sf F}_2$ vs. ${\sf F}_A={\sf F}_1$  																			& 99.27& 99.67  & 0.33  & 0.80 & 77.14 & 29.03 & 70.97 & 12.33 & 82.72 & 40.93  & 59.07  & 7.83          \\ \cline{2-14}
   & ${\sf F}_D={\sf F}_2$ vs. ${\sf F}_A={\sf F}_2$   																			& 99.52& 99.53  & 0.47  & 0.50 & 93.44 & 78.70 & 21.30 & 2.10  & 92.04 & 62.30  & 37.70  & 2.11       	\\ \hline
\multirow{3}{*}{${\sf ST}_A={\sf ST}_D$}& ${\sf F}_D={\sf F}_1$ vs. ${\sf F}_A={\sf F}_1$ & 
99.68& 99.44 & 0.56  & 0.20  & 96.92 & 96.32 & 3.68  & 2.89  & 95.73 & 92.03 & 7.97 	  & 3.27 \\ \cline{2-14}
\multirow{4}{*}{$={\tt full}$}   & ${\sf F}_D={\sf F}_1$ vs. ${\sf F}_A={\sf F}_2$  															& 99.27& 99.58  & 0.42  & 0.72 & 85.68 & 40.32 & 59.68 & 4.38  & 78.11 & 29.99  & 70.01  & 11.00          \\ \cline{2-14}
          & ${\sf F}_D={\sf F}_2$ vs. ${\sf F}_A={\sf F}_1$  																		& 99.60& 99.66  & 0.34  & 0.40 & 85.65 & 51.84 & 48.16 & 6.93  & 87.61 & 72.99  & 27.01  & 9.01          \\ \cline{2-14}
   & ${\sf F}_D={\sf F}_2$ vs. ${\sf F}_A={\sf F}_2$  & 99.68& 99.60 &  0.40  & 0.28 & 96.92 & 95.60 & 4.40 & 2.88	 & 95.73 & 90.09  & 9.91   &  2.83		\\ \hline
\end{tabular}
}
\vspace{-0.1in}
\end{table*}

\begin{table*}[!htbp]
 \centering
\caption{Proactive detection against adaptive attacks with ${\sf F}_D={\sf F}_A$.
For the baseline case $M_0(D_0)$, we have ACC = 99.68\%, TP =99.21\%, FN=0.79\%, FP=0.14\%.
\label{table:better-strategies}}
{\scriptsize
\begin{tabular}{|l|l|rrrr|rrrr|rrrr|}
\hline
\multirow{2}{*}{${\sf ST}_D$ vs. ${\sf ST}_A$} & \multirow{2}{*}{$M_{0\text{-}\gamma}(D_\alpha)$} & 
\multicolumn{4}{c|}{${\sf ST}_A={\tt parallel}$} & \multicolumn{4}{c|}{${\sf ST}_A={\tt sequential}$}  & \multicolumn{4}{c|}{${\sf ST}_A={\tt full}$} \\
\cline{3-14}
                      &           & ACC & TP & FN & FP & ACC  & TP & FN & FP  & ACC & TP & FN & FP  \\
\hline \hline
\multicolumn{14}{|c|}{Manipulation algorithm ${\sf F}_D={\sf F}_A={\sf F}_1$} \\
\hline
\multirow{2}{*}{${\sf ST}_D={\tt parallel}$}&
$M_{0\text{-}8}(D_1)$                         & 95.58& 92.03 & 7.97 & 3.62     & 94.25& 90.89 & 9.11 &  4.96    & 94.91   & 92.08 & 7.92 & 4.32        \\ \cline{2-14}
 &$M_{0\text{-}8}(D_9)$                           & 95.39& 92.00 & 8.00 & 3.83     & 92.38& 80.03 & 19.97 & 4.89    & 93.19   & 84.32 & 15.68 & 4.54        \\ \hline
\multirow{2}{*}{${\sf ST}_D={\tt sequential}$} &
$M_{0\text{-}8}(D_1)$ 	                        & 92.15& 74.22 & 25.78 & 3.93    & 93.44& 77.48 & 22.52 & 3.05    & 92.79   & 76.32 & 23.68 & 3.07        \\ \cline{2-14}
 &$M_{0\text{-}8}(D_9)$                              	& 89.20& 58.39 & 41.61 & 4.11    & 92.04& 59.33 & 30.67 & 2.99    & 88.42   & 57.89 & 42.11 & 3.91        \\ \hline
\multirow{2}{*}{${\sf ST}_D={\tt full}$}  &
$M_{0\text{-}8}(D_1)$                             & 96.24& 94.98 & 5.02 & 3.42     & 96.46& 94.99 & 5.01 & 3.15     & 96.92   & 96.32 & 3.68 & 2.89         \\ \cline{2-14}
 &$M_{0\text{-}8}(D_9)$                              & 94.73& 90.01 & 9.99 & 4.21     & 94.70& 90.03 & 9.97 & 4.23     & 95.73   & 92.03 & 7.97 & 3.27         \\  \hline
\hline
\multicolumn{14}{|c|}{Manipulation algorithm ${\sf F}_D={\sf F}_A={\sf F}_2$} \\
\hline
\multirow{2}{*}{${\sf ST}_D={\tt parallel}$}&
$M_{0\text{-}8}(D_1)$                                      	& 93.66	& 90.25 & 9.75 & 5.59    & 94.25 & 88.91 & 11.09 &  3.98    & 94.91  & 89.77 & 10.23 & 3.53            \\ \cline{2-14}
 &$M_{0\text{-}8}(D_9)$                                         & 96.17	& 92.77 & 7.23 & 3.08    & 92.38 & 77.89 & 22.11 & 4.32     & 93.19  & 81.32 & 18.68 & 3.38                 \\ \hline
\multirow{2}{*}{${\sf ST}_A={\tt sequential}$} &
$M_{0\text{-}8}(D_1)$ 	       	& 90.86	& 70.98 & 29.02 & 4.82   & 93.44 & 78.70 & 21.30 & 2.10     & 92.79  & 72.32 & 27.68 & 4.02                     \\ \cline{2-14}
 &$M_{0\text{-}8}(D_9)$                                         & 88.43	& 53.32 & 46.68 & 3.97   & 92.04 & 62.30 & 37.70 & 2.11     & 88.42  & 57.88 & 42.12 & 3.17            \\ \hline
\multirow{2}{*}{${\sf ST}_A={\tt full}$}  &
$M_{0\text{-}8}(D_1)$                         & 95.69	& 93.89 & 6.11 & 3.88    & 96.46 & 94.98 & 5.02 & 3.03      & 96.92  & 95.60 & 4.40 & 2.88             \\         \cline{2-14}
 &$M_{0\text{-}8}(D_9)$                               & 96.06	& 91.46 & 8.54 & 2.89    & 94.70 & 90.99 & 9.01 & 2.32      & 95.73  & 90.09 & 9.91 & 2.83           \\  \hline
\end{tabular}
}
\vspace{-0.1in}
\end{table*}

\begin{figure*}[!hbtp]
\centering
\subfigure[Fixed defender adaptation strategy against varying attacker adaptation strategies,
where both the attacker and the defender use manipulation algorithm ${\sf F}_1$.
We observe that the {\tt FULL} adaptation strategy leads to relatively better detection accuracy.\label{fig:gamma-alpha-f1}]{
\includegraphics[width=0.3\textwidth]{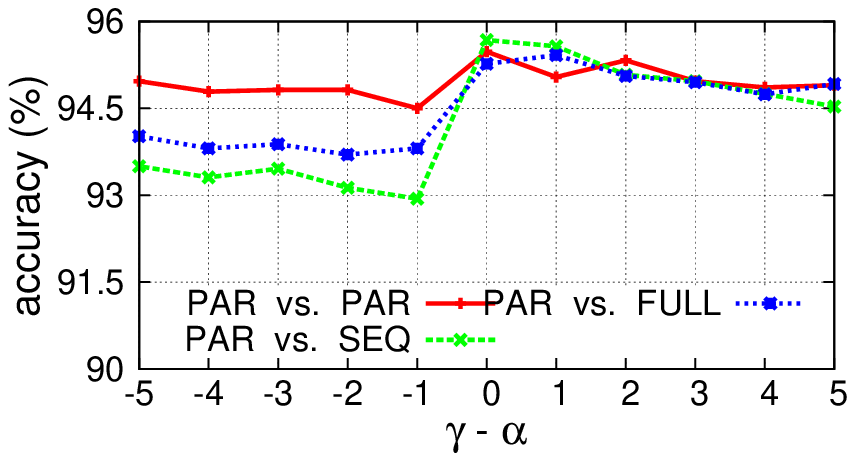}
\includegraphics[width=0.3\textwidth]{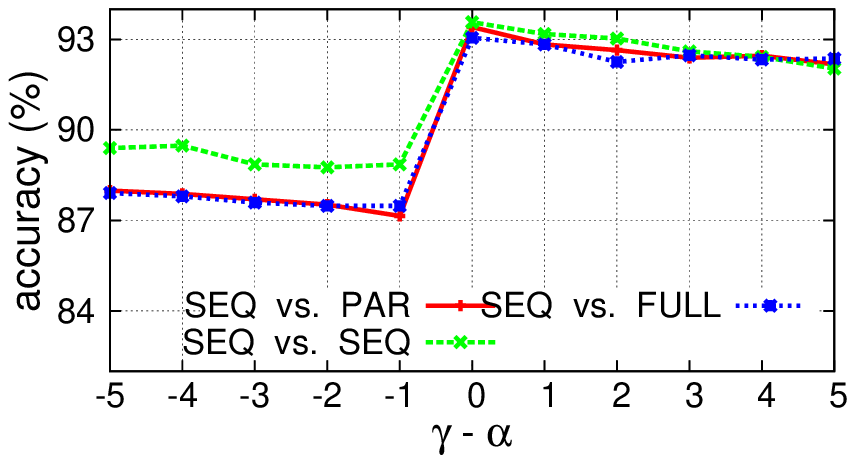}
\includegraphics[width=0.3\textwidth]{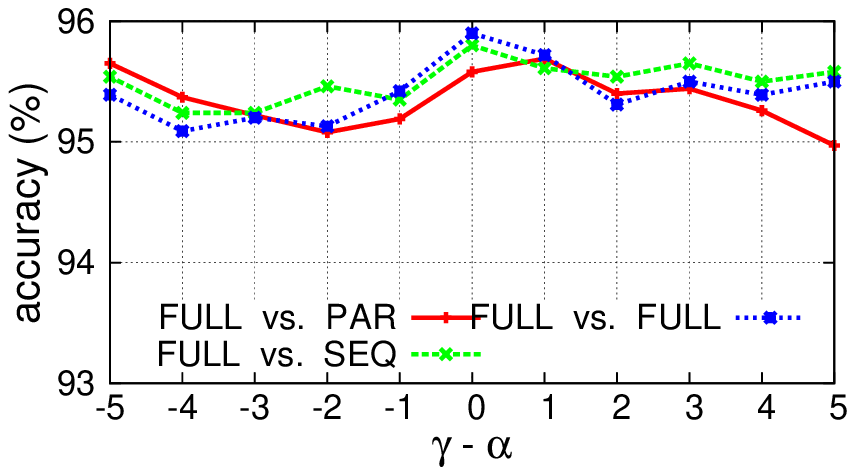}
}
\subfigure[Fixed defender adaptation strategy against varying attacker adaptation strategies,
where both the attacker and the defender use manipulation algorithm ${\sf F}_2$.
We observe that the {\tt FULL} adaptation strategy leads to relatively better detection accuracy.\label{fig:gamma-alpha-f2}]{
\includegraphics[width=0.3\textwidth]{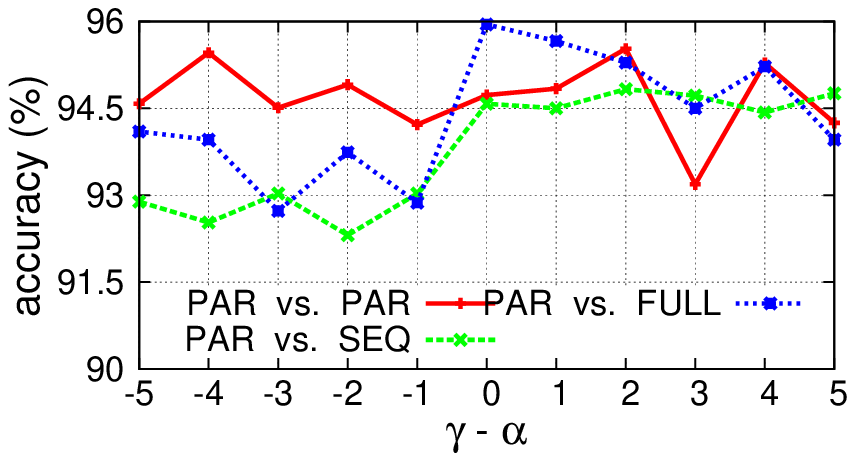}
\includegraphics[width=0.3\textwidth]{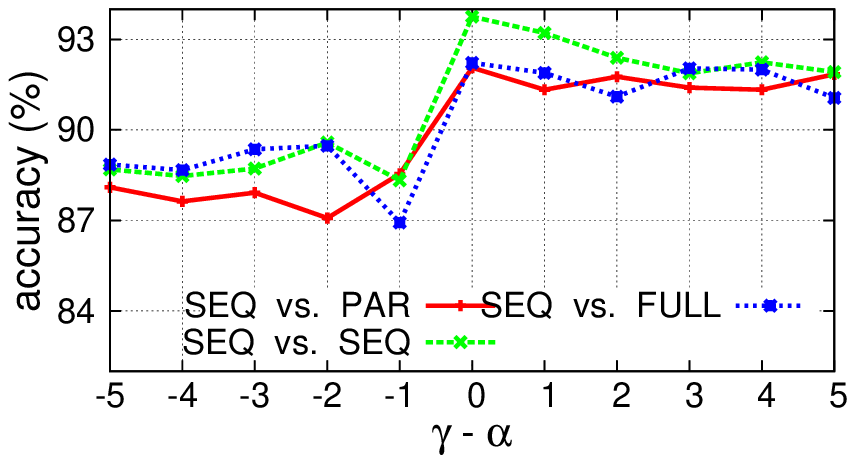}
\includegraphics[width=0.3\textwidth]{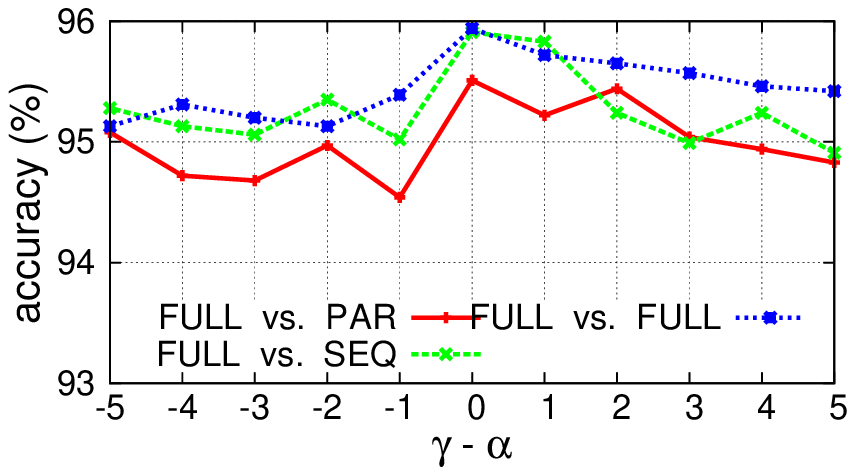}}
\caption{Impact of defender's proactiveness $\gamma$ vs. attacker's adaptiveness $\alpha$ on detection accuracy (average over the 40 days)
under various ``${\sf ST}_D\times {\sf ST}_A$'' combinations,
where $\alpha \in [0,8]$, $\gamma \in [0,9]$, ${\tt PAR}$, ${\tt SEQ}$ and ${\tt FULL}$ respectively stand for ${\tt parallel}$,
${\tt sequential}$ and ${\tt full}$ adaptation strategy, ``${\tt SEQ}$ vs. ${\tt APR}$" means ${\sf ST}_D={\tt sequential}$ and ${\sf ST}_A={\tt parallel}$ etc.
Note that $\gamma-\alpha=a$ is averaged over all possible combinations of $(\alpha,\gamma)$ as long as $\alpha \in [0,8]$ and $\gamma \in [0,9]$,
and that the detection accuracy is averaged over the 40 days. We observe that the detection accuracy in most cases there is a significant increase
in detection accuracy when the defender's proactiveness matches the attacker's adaptiveness.
\label{fig:gamma-alpha-f1-f2}}
\end{figure*}

Second, when ${\sf ST}_A={\sf ST}_D\neq0$,
it has a significant impact whether or not they use the same manipulation algorithm.
This phenomenon also can be explained by that the features that are often manipulated by ${\sf F}_1$
are very different from the features that are often manipulated by ${\sf F}_2$.
More specifically, when ${\sf F}_A={\sf F}_D$, the proactively learned classifiers $M_1^\dag,\ldots,M_\gamma^\dag$
would capture the ``maliciousness" information embedded in the manipulated data $D_\alpha$;
this would not be true when ${\sf F}_A\neq {\sf F}_D$.
Moreover, the ${\tt sequential}$ adaptation strategy appears to be more ``oblivious" than the other two strategies
in the sense that $D_\alpha$ preserves less information about $D_0$.
what adaptation strategy should the defender use to counter  ${\sf ST}_A={\tt sequential}$?
Table \ref{table:better-strategies} shows that the attacker does not have an obviously more effective counter ${\tt full}$ adaptation strategy.
This hints that the ${\tt full}$ strategy may be a kind of equilibrium strategy because
both attacker and defender have no significant gains by deviating from it. This inspires an important problem for future research:
{\bf Is the ${\tt full}$ adaptation strategy (or variant of it) an equilibrium strategy?}

Third, Table \ref{table:cross-and-aggregation} shows that when ${\sf ST}_D={\sf ST}_A$, the attacker can benefit by increasing
its adaptiveness $\alpha$. 
Table \ref{table:better-strategies} exhibits the same phenomenon when ${\sf ST}_D\neq {\sf ST}_A$.
In order to see the impact of defender's proactiveness as reflected by $\gamma$
against the defender's adaptiveness as reflected by $\alpha$,
we plot in Figure \ref{fig:gamma-alpha-f1-f2} how the detection accuracy with respect to $(\gamma-\alpha)$
under the condition ${\sf F}_D={\sf F}_A$ and under various ${\sf ST}_D\times {\sf ST}_A$ combinations.
We observe that roughly speaking, as $\gamma$ increases to exceed
the attacker's adaptiveness $\alpha$ (i.e., $\gamma$ changes from $\gamma<\alpha$ to $\gamma=\alpha$ to $\gamma>\alpha$),
the detection accuracy may have a significant increase at $\gamma-\alpha=0$.
Moreover, we observe that when ${\sf ST}_D={\tt full}$, $\gamma-\alpha$ has no significant impact on the detection accuracy.
This suggest that {\bf the defender should always use the ${\tt full}$ adaptation strategy to alleviate
the uncertainty about the attacker's adaptiveness $\alpha$}.

\section{Related Work}
\label{sec:related-work}

The problem of malicious websites detection has been an active research topic (e.g., \cite{driveby:prophiler:www:2011,Choi:2011:DMW:2002168.2002179,ucsd:evilseed}).
The dynamic detection approach has been investigated in \cite{js:www:10,conf/www/ZhangSSL11,Chen_ASIACCS11_WebPatrol,revolver:2013:usenix}.
The static detection approach has been investigated in \cite{dt:static2:2008,dt:static:2008,url.justin.kdd.09}.
The hybrid dynamic-static approach has been investigated in \cite{driveby:prophiler:www:2011,codaspy13:li,www:2013:hybrid}.
Loosely related to the problem of malicious websites detection are the detection of Phishing websites \cite{url.justin.kdd.09,url.ICML.09,phishing:framework:2007},
the detection of spams \cite{url:oakland:11,url:ndss:spam,url:ndss:ipcluster,url.justin.kdd.09,url.ICML.09},
the detection of suspicious URLs embedded in twitter message streams \cite{ndss12:twitter},
and the detection of browser-related attacks \cite{ndss13:postman,ndss13:chrome}.
However, none of these studies considered the issue of evasion by adaptive attackers.

The evasion attack is closely related to the problem of {\em adversarial machine learning},
where the attacker aims to evade an detection scheme that is derived from some machine learning
method \cite{berkeley:SecureML:Barreno,adversary:ndss:cmu}.
Perdisci et al. \cite{conf/icdm/PerdisciGL06}
investigated how to make the detection harder to evade.
Nelson et al. \cite{Berkeley:Evasion:Nelson} assumed the attacker has black-box access to the detection mechanism.
Dalvi et al. \cite{Washington:Nilesh} used Game Theoretic method to study this problem
in the setting of spam detection by assuming the attacker has access to the detection mechanism.
Our model actually gives attacker more freedom
because the attacker knows the data defender collected.

\section{Conclusion and Future Work}
\label{sec:conclusion}

We formulated a model of adaptive attacks by which the attacker can
manipulate the malicious websites to evade detection.
We also formulated a model of proactive defense against the adaptive attacks.
Experimental results based on a 40-day dataset showed that adaptive attacks can
evade non-proactive defense, but can be effectively countered by proactive defense.

In the full version of the present paper, we will address {\em correlation constraints} between features,
which is omitted due to space limitation.
This study also introduces a set of interesting research problems, including:
Are the same kinds of results/insights applicable to classifiers other than decision trees?
Is the ${\tt full}$ adaptation strategy indeed a kind of equilibrium strategy?
What is the optimal manipulation algorithm (if exists)?
How can we precisely characterize the evadability caused by adaptive attacks in this context?
What is the optimal time resolution at which the defender should proactively train its detection schemes (e.g., hour or day)?

\smallskip

\noindent{\bf Acknowledgement}. This work was supported in part by ARO Grant \#W911NF-13-1-0141.
%Any opinions, findings, and conclusions or recommendations expressed in this material are those of 
%the author(s) and do not necessarily reflect the views of the U.S. government.
This study was approved by IRB.

%\bibliographystyle{IEEEtran}
%\bibliography{li-xu}
%\bibliographystyle{abbrv}
%\bibliography{li-xu}

\end{document}